\documentclass[Afour,sageh,times]{sagej}

\usepackage{moreverb,url}

\usepackage[colorlinks,bookmarksopen,bookmarksnumbered,citecolor=red,urlcolor=red]{hyperref}
\usepackage{csquotes}
\usepackage{natbib}
\usepackage{listings}
\newcommand\BibTeX{{\rmfamily B\kern-.05em \textsc{i\kern-.025em b}\kern-.08em
T\kern-.1667em\lower.7ex\hbox{E}\kern-.125emX}}

\begin{document}

\runninghead{}

\title{Facilitating on-line opinion dynamics by mining expressions of causation. The case of climate change debates on \textit{The Guardian}}


\author{Tom Willaert\affilnum{1}, Sven Banisch\affilnum{2}, Paul Van Eecke\affilnum{1} and Katrien Beuls\affilnum{1}}

\affiliation{
\affilnum{1}VUB Artiﬁcial Intelligence Lab, Vrije Universiteit Brussel, Pleinlaan 2, 1050 Brussels, Belgium\\
\affilnum{2}Max Planck Institute for Mathematics in the Sciences, Inselstraße 22, Leipzig, Germany}



\begin{abstract}
News website comment sections are spaces where potentially conflicting opinions and beliefs are voiced. Addressing questions of how to study such cultural and societal conflicts through technological means, the present article critically examines possibilities and limitations of machine-guided exploration and potential facilitation of on-line opinion dynamics. These investigations are guided by a discussion of an experimental observatory for mining and analyzing opinions from climate change-related user comments on news articles from the TheGuardian.com. 
This observatory combines causal mapping methods with computational text analysis in order to mine beliefs and visualize opinion landscapes based on expressions of causation. By (1) introducing digital methods and open infrastructures for data exploration and analysis and (2) engaging in debates about the implications of such methods and infrastructures, notably in terms of the leap from opinion observation to debate facilitation, the article aims to make a practical and theoretical contribution to the study of opinion dynamics and conflict in new media environments. 
\end{abstract}

\maketitle

\section{Introduction}

\subsection{Background}

Over the past two decades, the rise of social media and the digitization of news and discussion platforms have radically transformed how individuals and groups create, process and share news and information. As Alan Rusbridger, former-editor-in-chief of the newspaper \textit{The Guardian} has it, these technologically-driven shifts in the ways people communicate, organize themselves and express their beliefs and opinions, have 

\begin{displayquote}
empower[ed] those that were never heard, creating a a new form of politics and turning traditional news corporations inside out. It is impossible to think of Donald Trump; of Brexit; of Bernie Sanders; of Podemos; of the growth of the far right in Europe; of the spasms of hope and violent despair in the Middle East and North Africa without thinking also of the total inversion of how news is created, shared and distributed. Much of it is liberating and and inspiring. Some of it is ugly and dark. And something - the centuries-old craft of journalism - is in danger of being lost  \citep[xx-xxi]{rusbridger2018breaking}. 
\end{displayquote}

Rusbridger's observation that the present media-ecology puts traditional notions of politics, journalism, trust and truth at stake is a widely shared one \citep[see for instance][]{mit2018politics, singer2018likewar, sunstein2018republic}. As such, it has sparked interdisciplinary investigations, diagnoses and ideas for remedies across the economical, socio-political, and technological spectrum, challenging our existing assumptions and epistemologies \citep[see][]{floridi2013philosophy, floridi2014fourth}. Among these lines of inquiry, particular strands of research from the computational social sciences are addressing pressing questions of how emerging technologies and digital methods might be operationalized to regain a grip on the dynamics that govern the flow of on-line news and its associated multitudes of voices, opinions and conflicts. Could the information circulating on on-line (social) news platforms for instance be mined to better understand and analyze the problems facing our contemporary society? Might such data mining and analysis help us to monitor the growing number of social conflicts and crises due to cultural differences and diverging world-views? And finally, would such an approach potentially facilitate early detection of conflicts and even ways to resolve them before they turn violent?

Answering these questions requires further advances in the study of cultural conflict based on digital media data. This includes the development of fine-grained representations of cultural conflict based on theoretically-informed text analysis, the integration of game-theoretical approaches to models of polarization and alignment, as well as the construction of accessible tools and media-monitoring observatories: platforms that foster insight into the complexities of social behaviour and opinion dynamics through automated computational analyses of (social) media data. Through an interdisciplinary approach, the present article aims to make both a practical and theoretical contribution to these aspects of the study of opinion dynamics and conflict in new media environments.

\subsection{Objective}

The objective of the present article is to critically examine possibilities and limitations of machine-guided exploration and potential facilitation of on-line opinion dynamics on the basis of an experimental data analytics pipeline or observatory for mining and analyzing climate change-related user comments from the news website of \textit{The Guardian} (TheGuardian.com). Combining insights from the social and political sciences with computational methods for the linguistic analysis of texts, this observatory provides a series of spatial (network) representations of the opinion landscapes on climate change on the basis of causation frames expressed in news website comments. This allows for the exploration of opinion spaces at different levels of detail and aggregation. 

Technical and theoretical questions related to the proposed method and infrastructure for the exploration and facilitation of debates will be discussed in three sections. The first section concerns notions of how to define what constitutes a belief or opinion and how these can be mined from texts. 
To this end, an approach based on the automated extraction of semantic frames expressing causation is proposed. The observatory thus builds on the theoretical premise that expressions of causation such as `global warming causes rises in sea levels' can be revelatory for a person or group's underlying belief systems.
Through a further technical description of the observatory's data-analytical components, section two of the paper deals with matters of spatially modelling the output of the semantic frame extractor and how this might be achieved without sacrificing nuances of meaning.  
The final section of the paper, then, discusses how insights gained from technologically observing opinion dynamics can inform conceptual modelling efforts and approaches to on-line opinion facilitation. As such, the paper brings into view and critically evaluates the fundamental conceptual leap from machine-guided observation to debate facilitation and intervention.

Through the case examples from \textit{The Guardian's} website and the theoretical discussions explored in these sections, the paper intends to make a twofold contribution to the fields of media studies, opinion dynamics and computational social science. 
Firstly, the paper introduces and chains together a number of data analytics components for social media monitoring (and facilitation) that were developed in the context of the \textless{}project name anonymized for review\textgreater{} infrastructure project. The \textless{}project name anonymized for review\textgreater{} infrastructure makes the components discussed in this paper available as open web services in order to foster reproducibility and further experimentation and development \textless{}infrastructure reference URL anonymized for review\textgreater{}. Secondly, and supplementing these technological and methodological gains, the paper addresses a number of theoretical, epistemological and ethical questions that are raised by experimental approaches to opinion exploration and facilitation. 
This notably includes methodological questions on the preservation of meaning through text and data mining, as well as the role of human interpretation, responsibility and incentivisation in observing and potentially facilitating opinion dynamics.

\subsection{Data: the communicative setting of TheGuardian.com}

In order to study on-line opinion dynamics and build the corresponding climate change opinion observatory discussed in this paper, a corpus of climate-change related news articles and news website comments was analyzed. Concretely, articles from the ‘climate change’ subsection from the news website of \textit{The Guardian} dated from 2009 up to April 2019 were processed, along with up to 200 comments and associated metadata for articles where commenting was enabled at the time of publication. The choice for studying opinion dynamics using data from \textit{The Guardian} is motivated by this news website's prominent position in the media landscape as well as its communicative setting, which is geared towards user engagement. Through this interaction with readers, the news platform embodies many of the recent shifts that characterize our present-day media ecology. 

TheGuardian.com is generally acknowledged to be one of the UK's leading online newspapers, with 8,2 million unique visitors per month as of May 2013 \citep{reid2013guardian}. The website consists of a core news site, as well as a range of subsections that allow for further classification and navigation of articles. Articles related to climate change can for instance be accessed by navigating through the `News' section, over the subsection `environment', to the subsubsection `climate change' \citep{guardian2019climate}. All articles on the website can be read free of charge, as \textit{The Guardian} relies on a business model that combines revenues from advertising, voluntary donations and paid subscriptions.

Apart from offering high-quality, independent journalism on a range of topics, a distinguishing characteristic of \textit{The Guardian} is its penchant for reader involvement and engagement. Adopting to the changing media landscape and appropriating business models that fit the transition from print to on-line news media, the Guardian has transformed itself into a platform that enables forms of citizen journalism, blogging, and welcomes readers comments on news articles \citep[for this transition, see for instance][chap. 11]{rusbridger2018breaking}. In order for a reader to comment on articles, it is required that a user account is made, which provides a user with a unique user name and a user profile page with a stable URL. According to the website's help pages, providing users with an identity that is consistently recognized by the community fosters proper on-line community behaviour \citep{guardian2019sign}. Registered users can post comments on content that is open to commenting, and these comments are moderated by a dedicated moderation team according to \textit{The Guardian's} community standards and participation guidelines \citep{guardian2019community}. In support of digital methods and innovative approaches to journalism and data mining, \textit{The Guardian} has launched an open API (application programming interface) through which developers can access different types of content \citep{guardian2019open}. It should be noted that at the moment of writing this article, readers' comments are not accessible through this API. For the scientific and educational purposes of this paper, comments were thus consulted using a dedicated scraper. 

\begin{figure}[ht]
 \centering
 \includegraphics[width=0.99\linewidth]{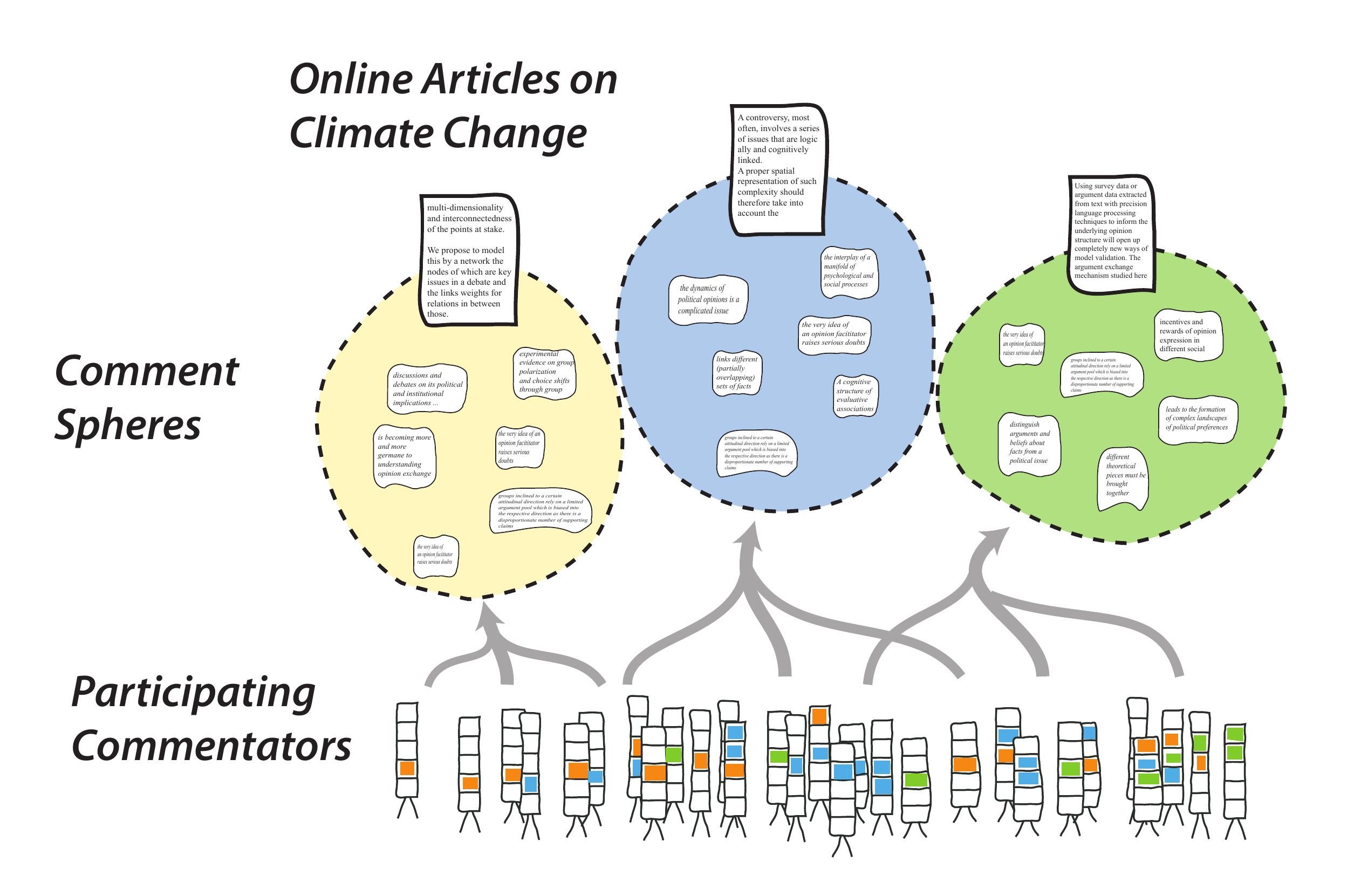}
 \caption{Communicative setting of many online newspaper sites. The newspaper publishes articles on different topics and users can comment on these articles and previous comments.}
 \label{fig:CommunicativeSetting}
 \end{figure}

Taking into account this community and technologically-driven orientation, the communicative setting of \textit{The Guardian} from which opinions are to be mined and the underlying belief system revealed, is defined by articles, participating commenters and comment spheres (that is, the actual comments aggregated by user, individual article or collection of articles) (see Figure \ref{fig:CommunicativeSetting}). 

In this setting, articles (and previous comments on those articles) can be commented on by participating commenters, each of which bring to the debate his or her own opinions or belief system. What this belief system might consists of can be inferred on a number of levels, with varying degrees of precision. On the most general level, a generic description of the profile of the average reader of \textit{The Guardian} can be informative. Such profiles have been compiled by market researchers with the purpose of informing advertisers about the demographic that might be reached through this news website (and other products carrying \textit{The Guardian's} brand). As of the writing of this article, the audience \textit{The Guardian} is presented to advertisers as a `progressive' audience: 

\begin{displayquote}
Living in a world of unprecedented societal change, with the public narratives around politics, gender, body image, sexuality and diet all being challenged. The Guardian is committed to reflecting the progressive agenda, and reaching the crowd that uphold those values. It’s helpful that we reach over half of progressives in the UK \citep{guardian2019advertising}. 
\end{displayquote}

A second, equally high-level indicator of the beliefs that might be present on the platform, are the links through which articles on climate change can be accessed. An article on climate change might for instance be consulted through the environment section of the news website, but also through the business section. Assuming that business interests might potentially be at odds with environmental concerns, it could be hypothesized that the particular comment sphere for that article consists of at least two potentially clashing frames of mind or belief systems. 

However, as will be expanded upon further in this article, truly capturing opinion dynamics requires a more systemic and fine-grained approach. The present article therefore proposes a method for harvesting opinions from the actual comment texts. The presupposition is thereby that comment spheres are marked by a diversity of potentially related opinions and beliefs. Opinions might for instance be connected through the reply structure that marks the comment section of an article, but this connection might also manifest itself on a semantic level (that is, the level of meaning or the actual contents of the comments). To capture this multidimensional, interconnected nature of the comment spheres, it is proposed to represent comment spheres as networks, where the nodes represent opinions and beliefs, and edges the relationships between these beliefs (see the spatial representation of beliefs \textit{infra}). The use of precision language tools to extract such beliefs and their mutual relationships, as will be explored in the following sections, can open up new pathways of  model validation and creation.

\section{Mining opinions and beliefs from texts}

In traditional experimental settings, survey techniques and associated statistical models provide researchers with established methods to gauge and analyze the opinions of a population. 
When studying opinion landscapes through on-line social media, however, harvesting beliefs from big textual data such as news website comments and developing or appropriating models for their analysis is a non-trivial task \citep[for an overview of methodological challenges facing computational social science and digital methods, see for instance][]{watts2013computational, rogers2013digital, rogers2019doing}.

In the present context, two challenges related to data-gathering and text mining need to be addressed: (1) defining what constitutes an expression of an opinion or belief, and (2) associating this definition with a pattern that might be extracted from texts. 
Recent scholarship in the fields of natural language processing (NLP) and argumentation mining has yielded a range of instruments and methods for the (automatic) identification of argumentative claims in texts \citep[see for instance][]{farzindar2017natural, stede2018argumentation}. Adding to these instruments and methods, the present article proposes an approach in which belief systems or opinions on climate change are accessed through expressions of causation. 

\subsection{Causal mapping methods and the climate change debate}

The climate change debate is often characterized by expressions of causation, that is, expressions linking a certain cause with a certain effect. Cultural or societal clashes on climate change might for instance concern diverging assessments of whether global warming is man-made or not \citep[for a sample of arguments in favour of or against anthropogenic global warming, see][]{procon2019}. Based on such examples, it can be stated that expressions of causation are closely associated with opinions or beliefs, and that as such, these expressions can be considered a valuable indicator for the range and diversity of the opinions and beliefs that constitute the climate change debate. The observatory under discussion therefore focuses on the extraction and analysis of linguistic patterns called causation frames. As will be further demonstrated in this section, the benefit of this causation-based approach is that it offers a systemic approach to opinion dynamics that comprises different layers of meaning, notably the cognitive or social meaningfulness of patterns on account of their being expressions of causation, as well as further lexical and semantic information that might be used for analysis and comparison. 

The study of expressions of causation as a method for accessing and assessing belief systems and opinions has been formalized and streamlined since the 1970s. Pioneered by political scientist Robert Axelrod and others, this causal mapping method (also referred to as `cognitive mapping') was introduced as a means of reconstructing and evaluating administrative and political decision-making processes, based on the principle that 

\begin{displayquote}
the notion of causation is vital to the process of evaluating alternatives. Regardless of philosophical difficulties involved in the meaning of causation, people do evaluate complex policy alternatives in terms of the consequences a particular choice would cause, and ultimately of what the sum of these effects would be. Indeed, such causal analysis is built into our language, and it would be very difficult for us to think completely in other terms, even if we tried \citep[5]{axelrod2016cognitive}.  
\end{displayquote}

Axelrod's causal mapping method comprises a set of conventions to graphically represent networks of causes and effects (the nodes in a network) as well as the qualitative aspects of this relation (the network’s directed edges, notably assertions of whether the causal linkage is positive or negative). These causes and effects are to be extracted from relevant sources by means of a series of heuristics and an encoding scheme (it should be noted that for this task Axelrod had human readers in mind). The graphs resulting from these efforts provide a structural overview of the relations among causal assertions (and thus beliefs): 

\begin{displayquote}
The basic elements of the proposed system are quite simple. The concepts a person uses are represented as \textit{points}, and the causal links between these concepts are represented as \textit{arrows} between these points. This gives a pictorial representation of the causal assertions of a person as a graph of points and arrows. This kind of representation of assertions as a graph will be called a \textit{cognitive map}. The policy alternatives, all of the various causes and effects, the goals, and the ultimate utility of the decision maker can all be thought of as concept variables, and represented as points in the cognitive map. The real power of this approach appears when a cognitive map is pictured in graph form; it is then relatively easy to see how each of the concepts and causal relationships relate to each other, and to see the overall structure of the whole set of portrayed assertions \citep[5]{axelrod2016cognitive}.  
\end{displayquote}

In order to construct these cognitive maps based on textual information, Margaret Tucker Wrightson provides a set of reading and coding rules for extracting cause concepts, linkages (relations) and effect concepts from expressions in the English language. The assertion `Our present topic is the militarism of Germany, which is maintaining a state of tension in the Baltic Area' might for instance be encoded as follows: `the militarism of Germany' (cause concept), /+/ (a positive relationship), `maintaining a state of tension in the Baltic area' (effect concept) \citep[296-297]{wrightson2016documentary}. Emphasizing the role of human interpretation, it is acknowledged that no strict set of rules can capture the entire spectrum of causal assertions:

\begin{displayquote}
The fact that the English language is as varied as those who use it makes the coder's task complex and difficult. No set of rules will completely solve the problems he or she might encounter. These rules, however, provide the coder with guidelines which, if conscientiously followed, will result in outcomes meeting social scientific standards of comparative validity and reliability \citep[332]{wrightson2016documentary}.
\end{displayquote}

To facilitate the task of encoders, the causal mapping method has gone through various iterations since its original inception, all the while preserving its original premises. Recent software packages have for instance been devised to support the data encoding and drawing process \citep[see for instance][]{laukkanen2016comparative}. As such, causal or cognitive mapping has become an established opinion and decision mining method within political science, business and management, and other domains. It has notably proven to be a valuable method for the study of recent societal and cultural conflicts. Thomas Homer-Dixon \textit{et al.} for instance rely on cognitive-affective maps created from survey data to analyze interpretations of the housing crisis in Germany, Israeli attitudes toward the Western Wall, and moderate \textit{versus} skeptical positions on climate change \citep{homer2014conceptual}. Similarly, Duncan Shaw \textit{et al.} venture to answer the question of `Why did Brexit happen?' by building causal maps of nine televised debates that were broadcast during the four weeks leading up to the Brexit referendum \citep{shaw2017did}. 

In order to appropriate the method of causal mapping to the study of on-line opinion dynamics, it needs to expanded from applications at the scale of human readers and relatively small corpora of archival documents and survey answers, to the realm of `big' textual data and larger quantities of information. This attuning of cognitive mapping methods to the large-scale processing of texts required for media monitoring necessarily involves a degree of automation, as will be explored in the next section. 

\subsection{Automated causation tracking with the Penelope semantic frame extractor}

As outlined in the previous section, causal mapping is based on the extraction of so-called cause concepts, (causal) relations, and effect concepts from texts. The complexity of each of these these concepts can range from the relatively simple (as illustrated by the easily-identifiable cause and effect relation in the example of `German militarism' cited earlier), to more complex assertions such as `The development of international cooperation in all fields across the ideological frontiers will gradually remove the hostility and fear that poison international relations', which contains two effect concepts (viz. `the hostility that poisons international relations' and `the fear that poisons international relations'). As such, this statement would have to be encoded as a double relationship \citep[297-298]{wrightson2016documentary}.

The coding guidelines in \citet{wrightson2016documentary} further reflect that extracting cause and effect concepts from texts is an operation that works on both the syntactical and semantic levels of assertions. This can be illustrated by means of the guidelines for analyzing the aforementioned causal assertion on German militarism: 

\begin{displayquote}
 1. The first step is the realization of the relationship. Does a subject affect an object? 2. Having recognized that it does, the isolation of the cause and effects concepts is the second step. As the sentence structure indicates, "the militarism of Germany" is the causal concept, because it is the initiator of the action, while the direct object clause, "a state of tension in the Baltic area," constitutes that which is somehow influenced, the effect concept \citep[296]{wrightson2016documentary}.  
\end{displayquote}

In the field of computational linguistics, from which the present paper borrows part of its methods, this procedure for extracting information related to causal assertions from texts can be considered an instance of an operation called semantic frame extraction \citep[for the concept of semantic frames, see][]{fillmore1982frame}. A semantic frame captures a coherent part of the meaning of a sentence in a structured way. As documented in the FrameNet project \citep{baker1998berkeley}, the \textsc{Causation} frame is defined as follows:

\begin{displayquote}
A Cause causes an Effect. Alternatively, an Actor, a participant of a (implicit) Cause, may stand in for the Cause. The entity Affected by the Causation may stand in for the overall Effect situation or event \citep{framenet2001causation}.  
\end{displayquote}

In a linguistic utterance such as a statement in a news website comment, the \textsc{Causation} frame can be evoked by a series of \textit{lexical units}, such as `cause', `bring on', etc. In the example `If such a small earthquake CAUSES problems, just imagine a big one!', the \textsc{Causation} frame is triggered by the verb `causes', which therefore is called the \textit{frame evoking element}. The \textsc{Cause} slot is filled by `a small earthquake', the \textsc{Effect} slot by `problems' \citep{framenet2001causation}. 

In order to automatically mine cause and effects concepts from the corpus of comments on \textit{The Guardian}, the present paper uses the Penelope semantic frame extractor: a tool that exploits the fact that semantic frames can be expressed as form-meaning mappings called constructions. Notably, frames were extracted from \textit{Guardian} comments by focusing on the following lexical units (verbs, prepositions and conjunctions), listed in FrameNet as frame evoking elements of the \textsc{Causation} frame: \textsc{Cause.v}, \textsc{Due to.prep}, \textsc{Because.c}, \textsc{Because of.prep}, \textsc{Give rise to.v}, \textsc{Lead to.v} or \textsc{Result in.v}. 

As illustrated by the following examples, the strings output by the semantic frame extractor adhere closely to the original utterance, preserving all of the the comments' causation frames real-world noisiness: 

\begin{lstlisting}
{
    "causalRelations": [
		{
			"utterance": "Has anyone totted up the extra pollution on London streets emanating from traffic jams caused by Extinction Rebellion ?",
			"cause": "extinction rebellion",
			"effect": "traffic jams"
		}
	]
}
\end{lstlisting}

The output of the semantic frame extractor as such is used as the input for the ensuing pipeline components in the climate change opinion observatory. The aim of a further analysis of these frames is to find patterns in the beliefs and opinions they express. As will be discussed in the following section, which focuses on applications and cases, maintaining semantic nuances in this further analytic process foregrounds the role of models and aggregation levels.

\section{Analyses and applications}

Based on the presupposition that relations between causation frames reveal beliefs, the output of the semantic frame extractor creates various opportunities for exploring opinion landscapes and empirically validating conceptual models for opinion dynamics. 

In general, any alignment of conceptual models and real-world data is an exercise in compromising, as the idealized, abstract nature of models is likely to be at odds with the messiness of the actual data. Finding such a compromise might for instance involve a reduction of the simplicity or elegance of the model, or, on the other hand, an increased aggregation (and thus reduced granularity) of the data. 

Addressing this challenge, the current section reflects on questions of data modelling, aggregation and meaning by exploring, through case examples, different spatial representations of opinion landscapes mined from the TheGuardian.com's comment sphere. These spatial renditions will be understood as network visualizations in which nodes represent argumentative statements (beliefs) and edges the degree of similarity between these statements. On the most general level, then, such a representation can consists of an overview of all the causes expressed in the corpus of climate change-related \textit{Guardian} comments. This type of visualization provides a birds-eye view of the entire opinion landscape as mined from the comment texts. In turn, such a general overview might elicit more fine-grained, micro-level investigations, in which a particular cause is singled out and its more specific associated effects are mapped. These macro and micro level overviews come with their own proper potential for theory building and evaluation, as well as distinct requirements for the depth or detail of meaning that needs to be represented. To get the most general sense of an opinion landscape one might for instance be more tolerant of abstract renditions of beliefs (e.g. by reducing statements to their most frequently used terms), but for more fine-grained analysis one requires more context and nuance (e.g. adhering as closely as possible to the original comment). 

\subsection{Aggregation}

As follows from the above, one of the most fundamental questions when building automated tools to observe opinion dynamics that potentially aim at advising means of debate facilitation concerns the level of meaning aggregation. 
A clear argumentative or causal association between, for instance, climate change and catastrophic events such as floods or hurricanes may become detectable by automatic causal frame tracking at the scale of large collections of articles where this association might appear statistically more often, but detection comes with great challenges when the aim is to classify certain sets of only a few statements in more free expression environments such as comment spheres.

In other words, the problem of meaning aggregation is closely related to issues of scale and aggregation over utterances. 
The more fine-grained the semantic resolution is, that is, the more specific the cause or effect is that one is interested in, the less probable it is to observe the same statement twice. 
Moreover, with every independent variable (such as time, different commenters or user groups, etc.), less data on which fine-grained opinion statements are to be detected is available. In the present case of parsed comments from TheGuardian.com, providing insights into the belief system of individual commenters, even if all their statements are aggregated over time, relies on a relatively small set of argumentative statements.
This relative sparseness is in part due to the fact that the scope of the semantic frame extractor is confined to the frame evoking elements listed earlier, thus omitting more implicit assertions of causation (i.e. expressions of causation that can only be derived from context and from reading between the lines).

Similarly, as will be explored in the ensuing paragraphs, matters of scale and aggregation determine the types of further linguistic analyses that can be performed on the output of the frame extractor. Within the field of computational linguistics, various techniques have been developed to represent the meaning of words as vectors that capture the contexts in which these words are typically used. Such analyses might reveal patterns of statistical significance, but it is also likely that in creating novel, numerical representations of the original utterances, the semantic structure of argumentatively linked beliefs is lost.  

In sum, developing opinion observatories and (potential) debate facilitators entails finding a trade-off, or, in fact, a middle way between macro- and micro-level analyses.
On the one hand, one needs to leverage automated analysis methods at the scale of larger collections to maximum advantage. 
But one also needs to integrate opportunities to interactively zoom into specific aspects of interest and provide more fine-grained information at these levels down to the actual statements. This interplay between macro- and micro-level analyses is explored in the case studies below.

\subsection{Spatial renditions of TheGuardian.com's opinion landscape}

The main purpose of the observatory under discussion is to provide insight into the belief structures that characterize the opinion landscape on climate change. For reasons outlined above, this raises questions of how to represent opinions and, correspondingly, determining which representation is most suited as the atomic unit of comparison between opinions. In general terms, the desired outcome of further processing of the output of the semantic frame extractor is a network representation in which similar cause or effect strings are displayed in close proximity to one another. A high-level description of the pipeline under discussion thus goes as follows. In a first step, it can be decided whether one wants to map cause statements or effect statements. Next, the selected statements are grouped per commenter (i.e. a list is made of all cause statements or effect statements per commenter). These statements are filtered in order to retain only nouns, adjectives and verbs (thereby also omitting frequently occurring verbs such as ‘to be’). The remaining words are then lemmatized, that is, reduced to their dictionary forms. This output is finally translated into a network representation, whereby nodes represent (aggregated) statements, and edges express the semantic relatedness between statements (based on a set overlap whereby the number of shared lemmata are counted). 

As illustrated by two spatial renditions that were created using this approach and visualized using the network analysis tool Gephi \citep{bastian2009gephi}, the labels assigned to these nodes (lemmata, full statements, or other) can be appropriated to the scope of the analysis.

\subsubsection{A macro-level overview: causes addressed in the climate change debate}

Suppose one wants to get a first idea about the scope and diversity of an opinion landscape, without any preconceived notions of this landscape's structure or composition. One way of doing this would be to map all of the causes that are mentioned in comments related to articles on climate change, that is, creating an overview of all the causes that have been retrieved by the frame extractor in a single representation. Such a representation would not immediately provide the granularity to state what the beliefs or opinions in the debates actually \textit{are}, but rather, it might inspire a sense of what those opinions might be \textit{about}, thus pointing towards potentially interesting phenomena that might warrant closer examination. 

\begin{figure*}[ht]
 \centering
 \includegraphics[width=0.99\linewidth]{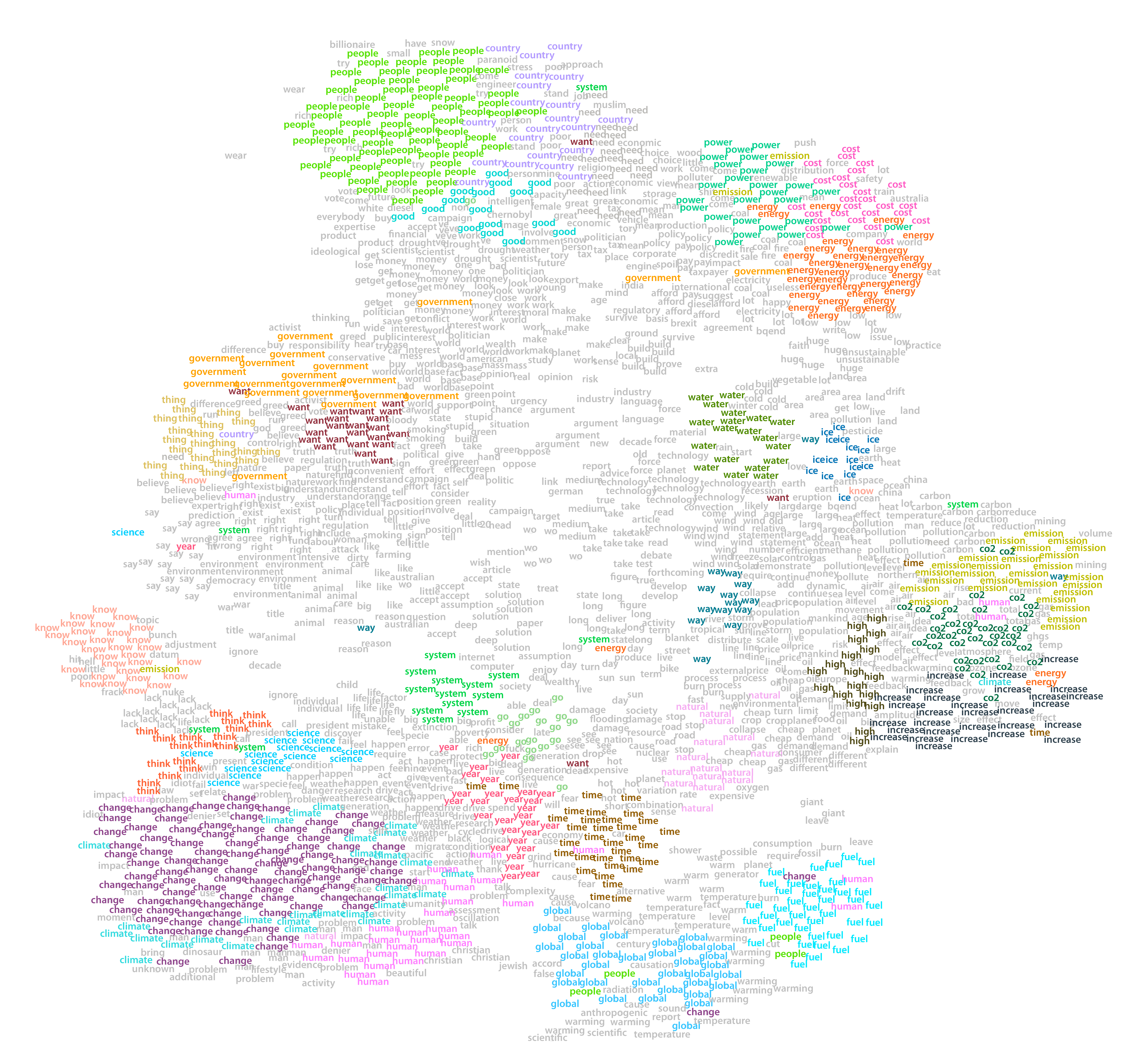}
 \caption{This is a global representation of the data produced by considering a 10 percent subsample of all the causes identified by the causation tracker on the set of comments. It treats statements as nodes of a network and two statements are linked if they share the same lemma (the number of shared lemmata corresponds to the link weight). In this analysis, only nouns, verbs and adjectives are considered (the text processing is done with spaCy \citep{honnibal2019spacy}). For this global view, each cause statement is labeled by that word within the statement that is most frequent in all the data. The visual output was created using the network exploration tool Gephi (0.92). The 2D layout is the result of the OpenOrd layout algorithm integrated in Gephi followed by the label adjustment tool to avoid too much overlap of labels.}
 \label{fig:GlobalStates}
 \end{figure*}

Figure \ref{fig:GlobalStates}, a high-level overview of the opinion landscape, reveals a number of areas to which opinions and beliefs might pertain. The top-left clusters in the diagram for instance reveal opinions about the role of people and countries, whereas on the right-hand side, we find a complementary cluster that might point to beliefs concerning the influence of high or increased CO2-emissions. In between, there is a cluster on power and energy sources, reflecting the energy debate's association to both issues of human responsibility and CO2 emissions. As such, the overview can already inspire, potentially at best, some very general hypotheses about the types of opinions that figure in the climate change debate.

\subsubsection{Micro-level investigations: opinions on nuclear power and global warming}

Based on the range of topics on which beliefs are expressed, a micro-level analysis can be conducted to reveal what those beliefs are and, for instance, whether they align or contradict each other. This can be achieved by singling out a cause of interest, and mapping out its associated effects.

\begin{figure}[ht]
 \centering
 \includegraphics[width=0.99\linewidth]{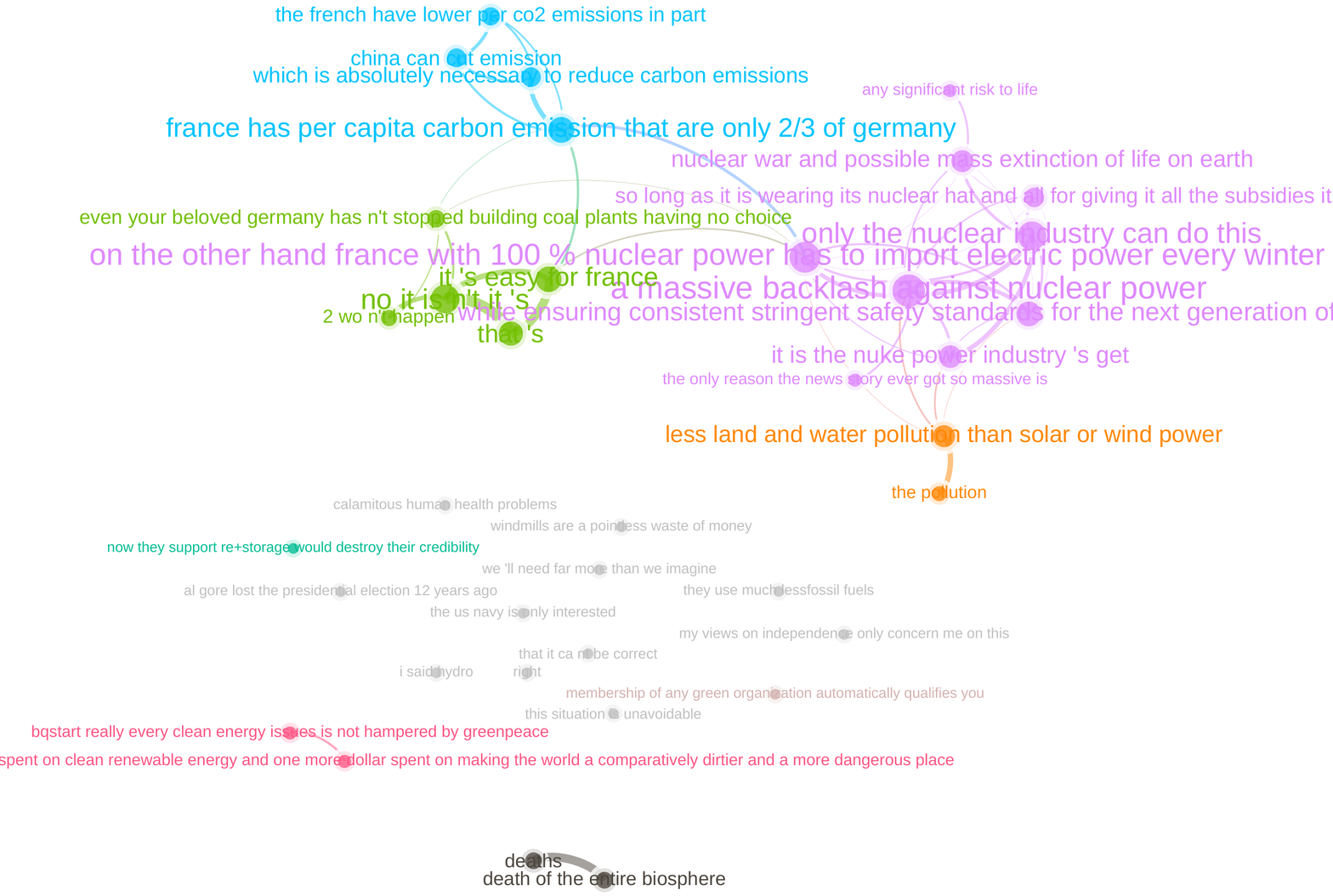}
 \caption{A detailed representation of effect statements associated with nuclear power. Clusters concern potential extinction or deaths, notions of cleanliness and pollution, and the reduction of CO2 levels in different countries. Labels represent the full output of the semantic frame extractor.}
 \label{fig:NuclearPower}
 \end{figure}

As revealed by the global overview of the climate change opinion landscape, a portion of the debate concerns power and energy sources. One topic with a particularly interesting role in this debate is nuclear power. Figure \ref{fig:NuclearPower} illustrates how a more detailed representation of opinions on this matter can be created by spatially representing all of the effects associated with causes containing the expression `nuclear power'. Again, similar beliefs (in terms of words used in the effects) are positioned closer to each other, thus facilitating the detection of clusters. Commenters on \textit{The Guardian} for instance express concerns about the deaths or extinction that might be caused by this energy resource. They also voice opinions on its cleanliness, whether or not it might decrease pollution or be its own source of pollution, and how it reduces CO2-emissions in different countries. 

\begin{figure*}[ht]
 \centering
 \includegraphics[width=0.99\linewidth]{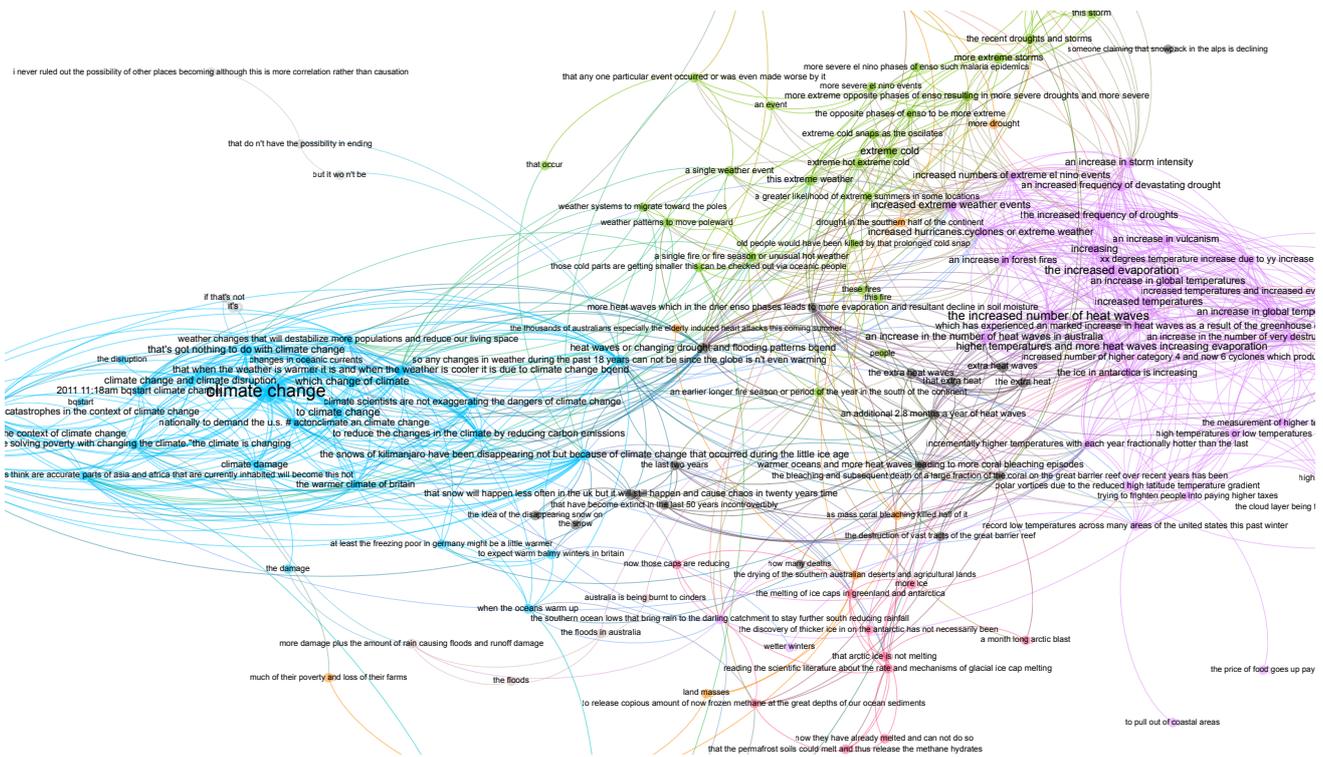}
 \caption{A detailed representation of the effects of global warming. This graph conveys the diversity of opinions, as well as emerging patterns. It can for instance be observed that certain opinions are clustered around the idea of `increases', notably in terms of evaporation, drought, heat waves, intensity of cyclones and storms, etc. An adjacent cluster is related to `extremes', such as extreme summers and weather events, but also extreme colds. Labels represent the full output of the semantic frame extractor.}
 \label{fig:GlobalWarming}
 \end{figure*}

Whereas the detailed opinion landscape on `nuclear power' is relatively limited in terms of the number of mined opinions, other topics might reveal more elaborate belief systems. This is for instance the case for the phenomenon of `global warming'. As shown in Figure \ref{fig:GlobalWarming}, opinions on global warming are clustered around the idea of `increases', notably in terms of evaporation, drought, heat waves, intensity of cyclones and storms, etc. An adjacent cluster is related to `extremes', such as extreme summers and weather events, but also extreme colds. 

\section{From opinion observation to debate facilitation}

The observatory introduced in the preceding paragraphs provides preliminary insights into the range and scope of the beliefs that figure in climate change debates on TheGuardian.com. The observatory as such takes a distinctly descriptive stance, and aims to satisfy, at least in part, the information needs of researchers, activists, journalists and other stakeholders whose main concern is to document, investigate and understand on-line opinion dynamics. However, in the current information sphere, which is marked by polarization, misinformation and a close entanglement with real-world conflicts, taking a mere descriptive or neutral stance might not serve every stakeholder's needs. Indeed, given the often skewed relations between power and information, questions arise as to how media observations might in turn be translated into (political, social or economic) action. Knowledge about opinion dynamics might for instance inform interventions that remedy polarization or disarm conflict. In other words, the construction of (social) media observatories unavoidably lifts questions about the possibilities, limitations and, especially, implications of the machine-guided and human-incentivized \textit{facilitation} of on-line discussions and debates. 

Addressing these questions, the present paragraph introduces and explores the concept of a debate \textit{facilitator}, that is, a device that extends the capabilities of the previously discussed observatory to also promote more interesting and constructive discussions. Concretely, we will conceptualize a device that reveals how the personal opinion landscapes of commenters relate to each other (in terms of overlap or lack thereof), and we will discuss what steps might potentially be taken on the basis of such representation to balance the debate. Geared towards possible interventions in the debate, such a device may thus go well beyond the observatory's objectives of making opinion processes and conflicts more transparent, which concomitantly raises a number of serious concerns that need to be acknowledged. 

On rather fundamental ground, tools that steer debates in one way or another may easily become manipulative and dangerous instruments in the hands of certain interest groups. Various aspects of our daily lives are for instance already implicitly guided by recommender systems, the purpose and impact of which can be rather opaque. For this reason, research efforts across disciplines are directed at scrutinizing and rendering such systems more transparent \citep{milano2019recommender}. Such scrutiny is particularly pressing in the context of interventions on on-line communication platforms, which have already been argued to enforce affective communication styles that feed rather than resolve conflict. The objectives behind any facilitation device should therefore be made maximally transparent and potential biases should be fully acknowledged at every level, from data ingest to the dissemination of results \citep[for a thorough discussion of challenges facing social media research in a post-truth era, see][]{rogers2018social}. More concretely, the endeavour of constructing opinion observatories and facilitators foregrounds matters of `openness' of data and tools, security, ensuring data quality and representative sampling, accounting for evolving data legislation and policy, building communities and trust, and envisioning beneficial implications. By documenting the development process for a potential facilitation device, the present paper aims to contribute to these on-going investigations and debates. Furthermore, every effort has been made to protect the identities of the commenters involved. In the words of media and technology visionary Jaron Lanier, developers and computational social scientists entering this space should remain fundamentally aware of the fact that `digital information is really just people in disguise' \citep[19]{lanier2013owns}.

With these reservations in mind, the proposed approach can be situated among ongoing efforts that lead from debate observation to facilitation. One such pathway, for instance, involves the construction of filters to detect hate speech, misinformation and other forms of expression that might render debates toxic \citep[see for instance][]{desmedt2018automatic, vanhee2018automatic}. 
Combined with community outreach, language-based filtering and detection tools have proven to raise awareness among social media users about the nature and potential implications of their on-line contributions \citep[see][]{projectgrey2019}. Similarly, advances can be expected from approaches that aim to extend the scope of analysis beyond descriptions of a present debate situation in order to model how a debate might evolve over time and how intentions of the participants could be included in such an analysis. 

Progress in any of these areas hinges on a further integration of real-world data in the modelling process, as well as a further socio-technical and media-theoretical investigation of how activity on social media platforms and technologies correlate to real-world conflicts. 
The remainder of this section therefore ventures to explore how conceptual argument communication models for polarization and alignment \citep{Banisch2018argument} might be reconciled with real-world data, and how such models might inform debate facilitation efforts.

\subsubsection{Debate facilitation through models of alignment and polarization}

As discussed in previous sections, news websites like TheGuardian.com establish a communicative settings in which agents (users, commenters) exchange arguments about different issues or topics. For those seeking to establish a healthy debate, it could thus be of interest to know how different users relate to each other in terms of their beliefs about a certain issue or topic (in this case climate change). Which beliefs are for instance shared by users and which ones are not? In other words, can we map patterns of alignment or polarization among users? 

\begin{figure*}[ht]
 \centering
 \includegraphics[width=0.99\linewidth]{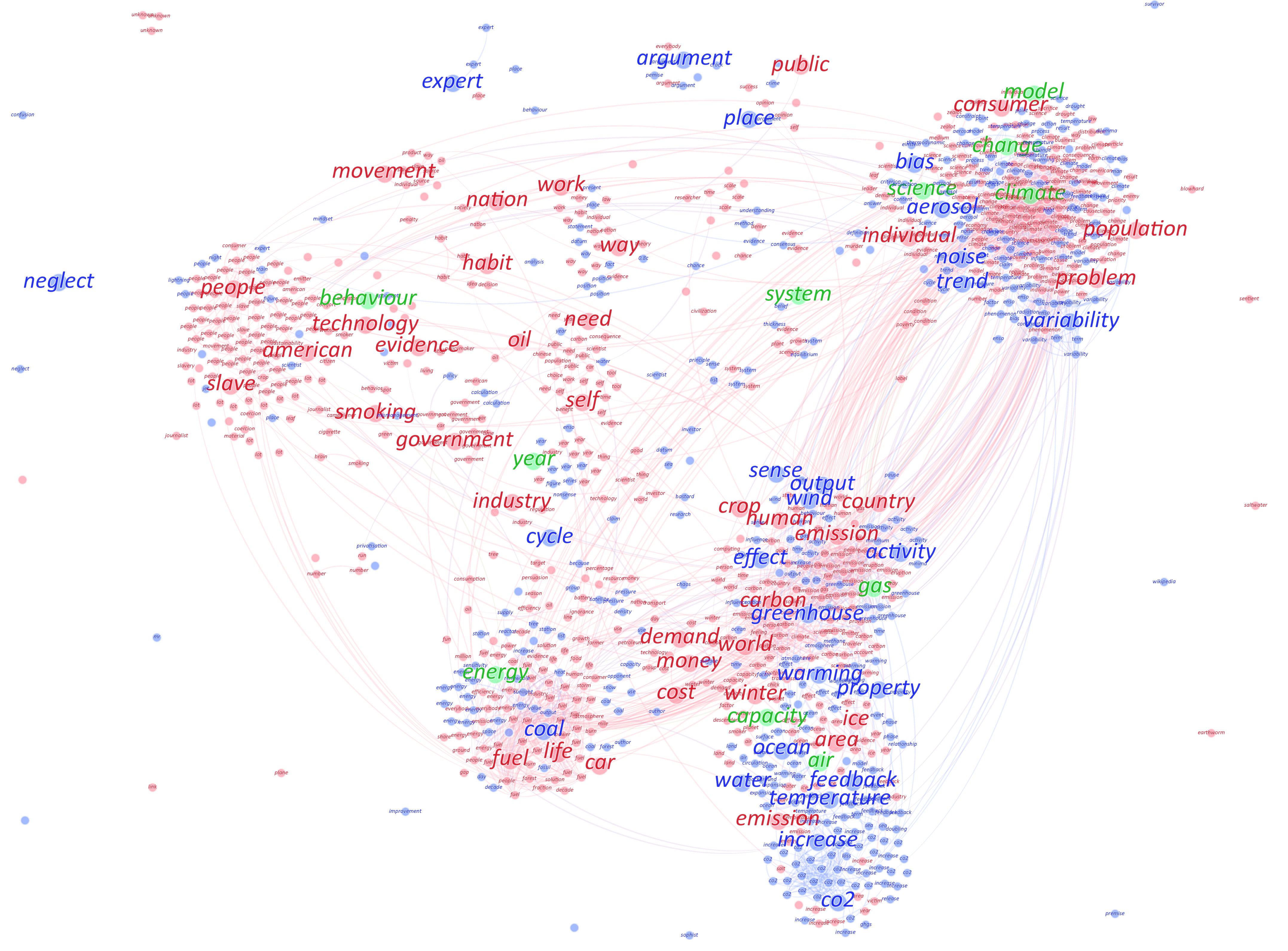}
 \caption{A representation of the opinion landscapes of two active commenters on TheGuardian.com. Statements by the first commenter are marked with a blue colour, opinions by the second commenter with a red colour. Overlapping statements are marked in green. The graph reveals that the commenters' beliefs are positioned most closely to each other on the most general aspects of the debate, whereas there is less relatedness on the social and more technical aspects of the discussion.}
 \label{fig:TwoUsersMostActiveMixed04AI}
 \end{figure*}

Figure \ref{fig:TwoUsersMostActiveMixed04AI} ventures to demonstrate how representations of opinion landscapes (generated using the methods outlined above) can be enriched with user information to answer such questions. Specifically, the graph represents the beliefs of two among the most active commenters in the corpus. The opinions of each user are marked using a colour coding scheme: red nodes represent the beliefs of the first user, blue nodes represent the beliefs of the second user. Nodes with a green colour represent beliefs that are shared by both users.

Taking into account again the factors of aggregation that were discussed in the previous section, Figure \ref{fig:TwoUsersMostActiveMixed04AI} supports some preliminary observations about the relationship between the two users in terms of their beliefs. Generally, given the fact that the graph concerns the two most active commenters on the website, it can be seen that the rendered opinion landscape is quite extensive. 
It is also clear that the belief systems of both users are not unrelated, as nodes of all colours can be found distributed throughout the graph. This is especially the case for the right-hand top cluster and right-hand bottom cluster of the graph, where green, red, and blue nodes are mixed. Since both users are discussing on articles on climate change, a degree of affinity between opinions or beliefs is to be expected. 

Upon closer examination, a number of disparities between the belief systems of the two commenters can be detected. Considering the left-hand top cluster and center of the graph, it becomes clear that exclusively the red commenter is using a selection of terms related to the economical and socio-political realm (e.g. `people', `american', `nation', `government') and industry (e.g. `fuel', `industry', `car', etc.). The blue commenter, on the other hand, exclusively engages in using a range of terms that could be deemed more technical and scientific in nature (e.g. `feedback', `property', `output', `trend', `variability', etc.). From the graph, it also follows that the blue commenter does not enter into the red commenter's `social' segments of the graph as frequently as the red commenter enters the more scientifically-oriented clusters of the graph (although in the latter cases the red commenter does not use the specific technical terminology of the blue commenter). The cluster where both beliefs mingle the most (and where overlap can be observed), is the top right cluster. This overlap is constituted by very general terms (e.g. `climate', `change', and `science'). 
In sum, the graph reveals that the commenters' beliefs are positioned most closely to each other on the most general aspects of the debate, whereas there is less relatedness on the social and more technical aspects of the debate. In this regard, the depicted situation seemingly evokes currently on-going debates about the role or responsibilities of the people or individuals \textit{versus} that of experts when it comes to climate change \citep[see for instance][]{catz2016climate, maki2019responsible, byskov2019climate}.

What forms of debate facilitation, then, could be based on these observations? 
And what kind of collective effects can be expected?
As follows from the above, beliefs expressed by the two commenters shown here (which are selected based on their active participation rather than actual engagement or dialogue with one another) are to some extent complementary, as the blue commenter, who displays a scientifically-oriented system of beliefs, does not readily engage with the social topics discussed by the red commenter. 
As such, the overall opinion landscape of the climate change could potentially be enriched with novel perspectives if the blue commenter was invited to engage in a debate about such topics as industry and government. Similarly, one could explore the possibility of providing explanatory tools or additional references on occasions where the debate takes a more technical turn. 

However, argument-based models of collective attitude formation \citep{Maes2013short,Banisch2018argument} also tell us to be cautious about such potential interventions. 
Following the theory underlying these models, different opinion groups prevailing during different periods of a debate will activate different argumentative associations.
Facilitating exchange between users with complementary arguments supporting similar opinions may enforce biased argument pools \citep{Sunstein2002law} and lead to increasing polarization at the collective level.
In the example considered here the two commenters agree on the general topic, but the analysis suggests that they might have different opinions about the adequate direction of specific climate change action.
A more fine--grained automatic detection of cognitive and evaluative associations between arguments and opinions is needed for a reliable use of models to predict what would come out of facilitating exchange between two specific users.
In this regard, computational approaches to the linguistic analysis of texts such as semantic frame extraction offer productive opportunities for empirically modelling opinion dynamics. Extraction of causation frames allows one to disentangle cause-effect relations between semantic units, which provides a productive step towards mapping and measuring structures of cognitive associations. These opportunities are to be explored by future work.

\section{Conclusion}

Ongoing transitions from a print-based media ecology to on-line news and discussion platforms have put traditional forms of news production and consumption at stake. Many challenges related to how information is currently produced and consumed come to a head in news website comment sections, which harbour the potential of providing new insights into how cultural conflicts emerge and evolve. On the basis of an observatory for analyzing climate change-related comments from TheGuardian.com, this article has critically examined possibilities and limitations of the machine-assisted exploration and possible facilitation of on-line opinion dynamics and debates.

Beyond technical and modelling pathways, this examination brings into view broader methodological and epistemological aspects of the use of digital methods to capture and study the flow of on-line information and opinions. Notably, the proposed approaches lift questions of computational analysis and interpretation that can be tied to an overarching tension between `distant' and `close reading' \citep{moretti2013distant}.  
In other words, monitoring on-line opinion dynamics means embracing the challenges and associated trade-offs that come with investigating large quantities of information through computational, text-analytical means, but doing this in such a way that nuance and meaning are not lost in the process. 

Establishing productive cross-overs between the level of opinions mined at scale (for instance through the lens of causation frames) and the detailed, closer looks at specific conversations, interactions and contexts depends on a series of preliminaries. One of these is the continued availability of high-quality, accessible data. As the current on-line media ecology is recovering from recent privacy-related scandals (e.g. Cambridge Analytica), such data for obvious reasons is not always easy to come by. In the same legal and ethical vein, reproducibility and transparency of models is crucial to the further development of analytical tools and methods. As the experiments discussed in this paper have revealed, a key factor in this undertaking are human faculties of interpretation. Just like the encoding schemes introduced by Axelrod and others before the wide-spread use of computational methods, present-day pipelines and tools foreground the role of human agents as the primary source of meaning attribution.

\begin{acks}
\textless{}This project has received funding from the European Union’s Horizon 2020 research and innovation programme under grant agreement No 732942 (Opinion Dynamics and Cultural Conflict in European Spaces -- www.\textsc{Odycceus}.eu).\textgreater{}
\end{acks}

\end{document}